\documentclass{article}
\usepackage{spconf,amsmath,graphicx}
\usepackage{flushend}
\usepackage{algorithm}
\usepackage{algorithmicx}
\usepackage{mathtools}
\usepackage{algpseudocode}
\usepackage{subfigure}
\usepackage{multirow}
\usepackage{longtable}
\usepackage{amsthm}
\usepackage{amssymb }
\usepackage{amsmath}
\usepackage{bm}
\usepackage{mathrsfs}
\usepackage{amsfonts}
\usepackage{longtable}
\usepackage{multirow}
\usepackage{cite}
\usepackage{xcolor}
\usepackage{graphicx}
\usepackage{enumitem}

\graphicspath{ {./images/} }

\def\MM{\text{TH-EMG}}
\def\p{p}
\def\i{i}
\def\x{{\bm{x}}}
\def\xp{\x^{\p}}
\def\E{\bm{E}}
\def\z{\bm{z}}

\def\t{t}
\def\c{\bm{c}}
\def\N{N}
\def\T{T}
\def\Q{Q}
\def\K{K}
\def\D{D}
\def\s{\bm{s}}
\def\V{V}
\def\d{d}
\def\l{l}
\def\L{L}
\def\j{j}

\def\S{\bm{S}}

\def\V{\bm{V}}
\def\Q{\bm{Q}}
\def\K{\bm{K}}
\def\Z{\bm{Z}}

\def\E{\bm{E}}
\def\X{\bm{X}}
\def\N{N}
\def\p{p}

\def\d{d}
\def\w{w}

\def\l{l}
\def\h{h}
\def\M{M}
\def\L{L}
\def\k{k}
\def\C{\bm{C}}
\def\H{\bm{H}}
\def\MM{\text{HYDRA-HGR}}

\title{$\MM$: A Hybrid Transformer-based Architecture for 
Fusion of Macroscopic and Microscopic Neural Drive Information}

\ninept

\name{Mansooreh Montazerin$^\dag$, Elahe Rahimian$^{\ddag}$, Farnoosh Naderkhani$^\ddag$, S. Farokh Atashzar$^{\dag\dag}$\vspace{-.2in}}
\address{\textit{Hamid Alinejad-Rokny$^{\ddag\ddag}$, Arash Mohammadi$^\ddag$}\\
$^\dag$~Electrical and Computer Engineering, Concordia University, Montreal, Canada\\
$^\ddag$~Concordia Institute of Information Systems Engineering (CIISE), Montreal, Canada\\
$^{\dag\dag}$ New York University (NYU), New York, 10003, NY, USA\\
$^{\ddag\ddag}$ The Graduate School of Biomedical Engineering, UNSW SYDNEY, Sydney, NSW, 2052, Australia
}

\begin{document}

\maketitle

\begin{abstract}
Development of advance surface Electromyogram (sEMG)-based Human-Machine Interface (HMI) systems is of paramount importance to pave the way towards emergence of futuristic Cyber-Physical-Human (CPH) worlds. In this context, the main focus of recent literature was on development of different Deep Neural Network (DNN)-based architectures that perform Hand Gesture Recognition (HGR) at a macroscopic level (i.e., directly from sEMG signals). At the same time, advancements in acquisition of High-Density sEMG signals (HD-sEMG) have resulted in a surge of significant interest on sEMG decomposition techniques to extract microscopic neural drive information.
However, due to complexities of  sEMG decomposition and added computational overhead, HGR at microscopic level is less explored than its aforementioned DNN-based counterparts.
 In this regard, we propose the $\MM$ framework, which is a hybrid model that simultaneously extracts a set of temporal and spatial features through its two independent Vision Transformer (ViT)-based parallel architectures (the so called Macro and Micro paths). The Macro Path is trained directly on the pre-processed HD-sEMG signals, while the Micro path is fed with the p-to-p values of the extracted Motor Unit Action Potentials (MUAPs) of each source. Extracted features at macroscopic and microscopic levels are then coupled via a Fully Connected (FC) fusion layer. We evaluate the proposed hybrid $\MM$ framework through a recently released HD-sEMG dataset, and show that it significantly outperforms its stand-alone counterparts. The proposed $\MM$ framework achieves average accuracy of $94.86$\% for the $250$ ms window size, which is $5.52$ \% and $8.22$ \% higher than that of the Macro and Micro paths, respectively.

%
\end{abstract}
\maketitle

\vspace{-.15in}
\section{Introduction} \label{sec:intro}
\vspace{-.1in}
In recent years, development of prosthetic Human-Machine Interface (HMI) systems~\cite{dario_muap} has greatly improved lives of those suffering from amputated limb(s) or neuromuscular disorders. Generally speaking, recent HMI systems~\cite{hmi} are mainly controlled with a learning model developed based on Biological Signal Processing (BSP) algorithms. Learning-based algorithms developed for potential use in prosthetic HMI devices are primarily comprised of advanced Machine Learning (ML) models~\cite{adv_ml} or Deep Neural Networks (DNNs)~\cite{dnn} that perform Hand Gesture Recognition (HGR) via surface Electromyogram (sEMG) signals. To secure high performance and proper control in HGR tasks, attention is directed to DNNs and/or hybrid architectures because of their proven capabilities in tackling challenging real-life problems~\cite{elahe_icassp,soheiltra,compare}. Such  models, however, are still prone to several major challenges such as high latency, large training times, undue complexity, and difficulty in finding the discriminative patterns in sEMG signals of different hand gestures. Accordingly, designing an accurate HGR method that can be effectively adopted in an HMI system addressing the above-mentioned issues is the centerpiece of today's gesture recognition-based research works.

\noindent
\textbf{Literature Review:} sEMG signals measure the electrical activities of the underlying motor units in limb muscles and are collected non-invasively from the electrodes placed on skin surface~\cite{semg}. In particular, High Density sEMG (HD-sEMG) signals are acquired through a two-dimensional (2D) grid with a large number of closely-located electrodes~\cite{high_semg}, capturing both temporal and spatial information of muscle activities. HD-sEMG acquisition, therefore, provides superior spatial resolution of the neuromuscular system in comparison to its sparse acquisition counterpart. This has inspired targeted focus on development of DNN-based HGR methods based on HD-sEMG signals~\cite{embc, instant, hybrid, hd_farokh}. Broadly speaking, HD-sEMG-based BSP approaches can be classified into the following two main categories:
\begin{itemize}[leftmargin=*]
\vspace{-.05in}
    \item [(i)] \textit{Raw HD-sEMG Processing for HGR:} Algorithms belonging to this category directly use raw HD-sEMG signals for the task of HGR. In this context, e.g., Reference~\cite{instant} performed instantaneous training of a Convolutional Neural Network (CNN) using a 2D image of a single time measurement. In~\cite{hybrid}, Recurrent Neural Networks (RNNs) are combined with CNNs to create a hybrid attention-based~\cite{attention} CNN-RNN architecture, which has improved HGR performance due to joint incorporation of spatial and temporal features of HD-sEMG signals. Sun \textit{et~al.}~\cite{hd_farokh} introduced a network of dilated Long Short-Term Memories (LSTMs) to classify hand gestures from the transient phase of HD-sEMG signals.

   \item[(ii)] \textit{HD-sEMG Decomposition:} The focus here is on HD-sEMG decomposition to extract microscopic neural drive information. HD-sEMG signals have encouraged emergence of sEMG decomposition algorithms in the last decade~\cite{bss} as they provide a significantly high-resolution 2D image of Motor Unit (MU) activities in each time point. sEMG decomposition refers to a set of Blind Source Separation (BSS)~\cite{bss_survey} methods that extract discharge timings of motor neuron action potentials from raw HD-sEMG data. Single motor neuron action potentials are summed to form Motor Unit Action Potentials (MUAPs) that convert neural drive information to hand movements~\cite{neural}. Motor unit discharge timings, also known as Motor Unit Spike Trains (MUSTs), represent sparse estimations of the MU activation times with the same sampling frequency and time interval as the raw HD-sEMG signals~\cite{farokh_dec}. Extracted MUSTs are used in several domains such as identification of motor neuron diseases~\cite{dec2}, analysis of neuromuscular  conditions~\cite{dec1}, and myoelectric pattern recognition~\cite{zhao_muap}.
   \vspace{-.05in}
\end{itemize}
A third category can be identified when the extracted MUSTs in Category (ii) are used for HGR at microscopic level. HD-sEMG signals are modelled as a spatio-temporal convolution of MUSTs, which provide an exact physiological description of how each hand movement is encoded at neurospinal level~\cite{nature_dario}. Thus, MUSTs are of trustworthy and discernible information on the generation details of different hand gestures, which leads to adoption of another group of HGR algorithms that accept MUSTs~\cite{relia} as their input. Nevertheless, due to complexities of the decomposition stage and added computational overhead, microscopic level HGR using MUST is less explored than models of Category (i), which use HD-sEMG signals at a macroscopic level.
There are, however, some promising works~\cite{zhao_muap,dario_muap,extracting} in which MUSTs carrying microscopic neural drive information are exploited for HGR instead of directly using raw sEMG signals. To discover a direct connection between different hand gestures and extracted MUSTs, these methods have suggested estimating MUAPs of the identified sources and extracting a set of useful features from MUAPs that are unique for each hand gesture. For instance, in~\cite{zhao_muap}, the peak-to-peak (p-to-p) values of MUAPs are calculated for each MU and each electrode channel separately and a 2D image of MUAP p-to-p are constructed for all the channels of a single MU. Afterwards, this 2D image is fed to a CNN architecture and its performance is compared to that of traditional ML methods. In short, using HD-sEMG decomposition results for HGR is still in its infancy, and in this paper, we aim to further advance this domain.

\vspace{.025in}
\noindent
\textbf{Contributions:} In this paper, for the first time to the best of our knowledge, we introduce the idea of integrating the macroscopic and microscopic neural drive information obtained from HD-sEMG data into a hybrid framework for HGR. As indicated in~\cite{elahe_icassp,embc}, Vision Transformers (ViTs)~\cite{ViT} are proved to surpass the most popular DNN architectures like CNNs and hybrid CNN-RNN models in HGR tasks, using either sparse or HD-sEMG datasets. Being motivated by this fact, here, we propose a framework that consists of two independent ViT-based architectures that work on the basis of the attention mechanism~\cite{embc}. While one of these ViTs is trained directly on the preprocessed HD-sEMG data, the other one is fed with the p-to-p values of the extracted MUAPs of each source. Thereby, a set of temporal and spatial features of various hand movements are extracted from HD-sEMG signals and MUAPs, which are then fused via a Fully Connected (FC) fusion layer for final classification. We show that by extracting a very small number of underlying sources (maximum of $7$ sources in this case), the proposed method improves the overall classification accuracy compared to its counterparts. In brief, contributions of the paper are as follows:
\vspace{-0.06in}

\begin{itemize}[leftmargin=*]
\item Introducing, for the first time to the best of our knowledge, the idea of integrating macroscopic and microscopic neural drive information through a hybrid DNN framework for HGR.
\vspace{-0.05in}
\item Simultaneous utilization of raw HD-sEMG signals and extracted MUSTs from HD-sEMG decomposition via parallel construction of two ViT-based architectures.
\end{itemize}
%

\vspace{-0.2in}
\section{Materials and Method}\label{sec:Int}
\vspace{-0.1in}
\setlength{\textfloatsep}{0pt}
\begin{figure}[t!]
\centering
\includegraphics[scale=0.3]{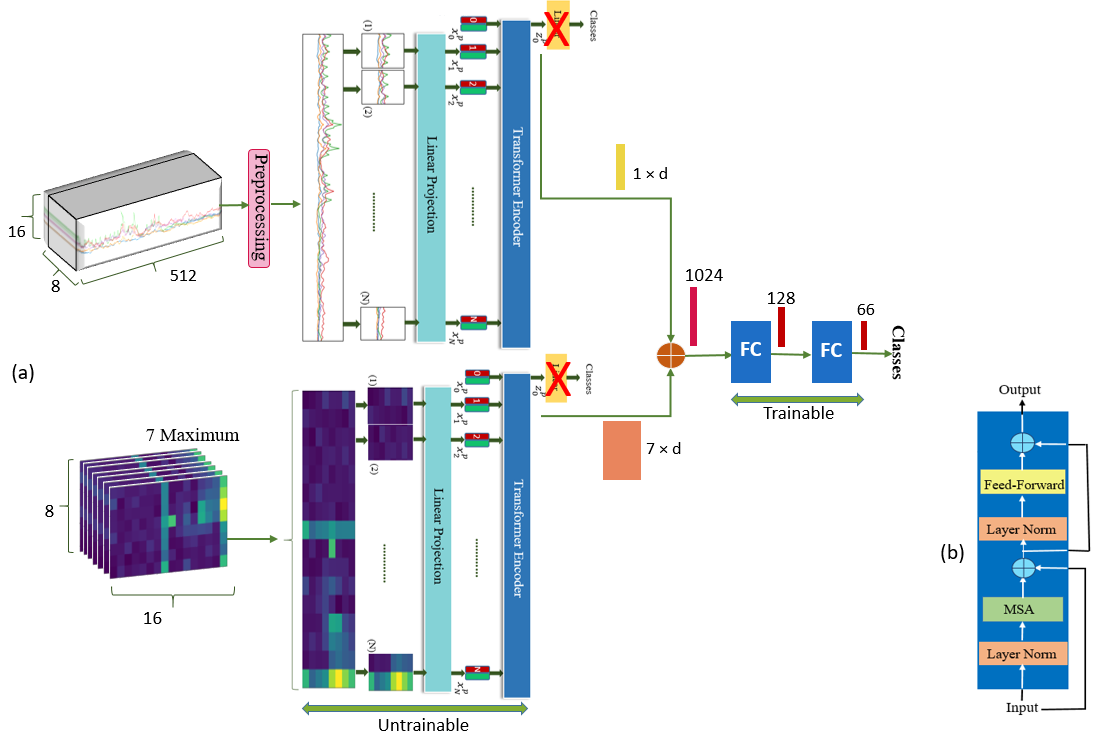}
\vspace{-0.45cm}
\caption{\footnotesize The proposed $\MM$: (a) The ViT-based models in  the Macro and Micro paths are trained based on 3D, HD-sEMG and 2D, p-to-p MUAP images, respectively. The final Micro and Macro class tokens are concatenated and converted to a $1,024$-dimensional feature vector, which is fed to a series of FC layers. (b) The transformer encoder.}
\label{hybrid}
\end{figure}
In this section, first, we introduce the dataset used to develop the proposed hybrid $\MM$ . Afterwards, we briefly describe the transformer architecture as its building block.

\vspace{.025in}
\noindent
\textbf{2.1. Dataset}

\noindent
The utilized dataset~\cite{data} in this study is a HD-sEMG collection of $20$ non-disabled participants that performed $65$ isometric hand movements in $5$ repetitions. One of the movements is performed twice, therefore, we have $66$ gestures in total. The dataset is a set of $1$-, $2$- and multiple-degree of freedom hand gestures. There are two high-density grids, each with $64$ ($8\times 8$) electrodes and an inter-electrode distance of $10$mm, placed on the flexor and extensor muscles to record the sEMG signals with the sampling frequency of $2,048$ Hz (Quattrocento, OT Bioelettronica, Torino, Italy).

\vspace{.05in}
\noindent
\textbf{2.2. Transformer Architecture}

\noindent
The proposed framework is originated from the ViT architecture in which the attention mechanism is deployed to find the similarities among different parts of the input data. In a ViT, the input, which is normally a three-dimensional (3D) image, is first divided into several patches and the sequence of patches are then fed as the main input. In the next stage, the 3D input patches are first flattened and then go through a patch embedding block, which converts each small flattened patch to an embedded vector to make them clearly understandable by the transformer network. This process is referred to as the linear projection of patches. Moreover, to encapsulate the learned information obtained during the training phase in a single vector, a zero vector (called the class token) is concatenated to the embedded patches similar to the BERT architecture~\cite{bert}. In this way, the final classification decision is made only based on the class token, which is attributed as the network's placeholder that prevents making decisions that are biased towards any of the patches as such reduces the number of trainable parameters in the model. ViTs, in contrast to RNNs, receive their input data in parallel, which makes it difficult for the network to identify the correct order in time-series data. Therefore, a trainable positional embedding vector is added to all the input patches, helping the network to understand their position more efficiently. These steps result in matrix $\Z_0$ as the input to the transformer encoder, which is given by
\begin{eqnarray}
\Z_0 = [\xp_0; \xp_1\E; \xp_2\E;\dots; \xp_\N\E] + \E^{pos},
\end{eqnarray}
where, $N$ is the number of patches, $\xp_\i$ represents the flattened patches that are multiplied by an embedding matrix $\E$, prepended with an all-zero class token and eventually added to a positional embedding matrix $\E^{pos}$. Note that all the embedded patches before and after going to the patch embedding block are of size $d$, which is the embedding dimension of the model that remains constant throughout all the training steps. Subsequently, $\Z_0$ is fed to the transformer encoder that contains the Multiheaded Self-Attention (MSA) mechanism in which the patches are divided into several parallel layers (heads). The scaled dot-product attention is distinctively found in each head. The corresponding results for each head are given to a feed forward layer and then, joined together to form the final decision. The attention metric is calculated based on different $d$-dimensional Queries ($\Q$), Keys ($\K$) and Values ($\V$) as follows
\vspace{-.075in}
\begin{eqnarray}
\text{$Attention(Q,K,V)$} = \text{$Softmax$}(\frac{\Q\K^T}{\sqrt{\d}})\V.
\end{eqnarray}
\vspace{-.05in}
The output matrix of the transformer encoder is then given by
\begin{eqnarray}
&&\!\!\!\!\!\!\!\!\!\!\!\!\!\!\!\!\bm{\mathcal{Z}}_0 = [\z^{\p}_{0}; \z^{\p}_{1}; \dots; \z^{\p}_{N}]= \textit{Feed-Forward}(LN(\bm{\mathbb{Z}}_0)) + \bm{\mathbb{Z}}_0,\\
&&\!\!\!\!\!\!\!\!\!\!\!\!\!\!\!\!\text{where } \quad\quad \bm{\mathbb{Z}}_0 = MSA(LN(\Z_0)) + \Z_0,\label{eq:MSA}
\end{eqnarray}
and $\z^{\p}_{\i}$ is the final output for the $\i^{th}$ patch, and $LN$ stands for the layer normalization block. In the last stage, the output class token ($\z^{\p}_{0}$) is forwarded to a linear layer to form the classification target.
This completes an overview of the preliminary material, next, we present the proposed $\MM$ framework.

\vspace{-0.15in}
\section{Proposed $\MM$ Architecture}\label{sec:MM}
\vspace{-0.1in}
In this section, we present the proposed hybrid ViT-based architecture that couples HD-sEMG signals and their extracted MUAPs. The $\MM$ framework simultaneously extracts a set of temporal and spatial features through its two independent ViT-based parallel paths. The \textit{Macro Path} is trained directly on the pre-processed HD-sEMG signals, while the \textit{Micro path} is fed with the p-to-p values of the extracted MUAPs of each source. A fusion path, structured in series to the parallel ones and consisting of FC layers, then combines extracted temporal and spatial features for final classification.

\vspace{.05in}
\noindent
\textbf{3.1. Macro Path: Extraction of Macroscopic Information}

\noindent
In this sub-section, we present the Macro Path of the $\MM$ framework, which directly learns from raw HD-sEMG signals. As HD-sEMG signals are of $3$ dimensions, i.e., one dimension in time and two dimensions in space, they can directly be fed to a ViT that is designed for 3D images. The raw HD-sEMG signals, however, should be first pre-processed to be entirely compatible with the ViT's input format. First, the positive envelop of the data is found via a low-pass butterworth filter (at $1$Hz), which is applied to signals of each electrode independently~\cite{instant,embc}. Then, the filtered signals are normalized through the $\mu$-law algorithm~\cite{xception}, which increases the selectivity power of the model. Following that, each HD-sEMG repetition is split into windows of size $512$ ($250$ ms) with a skip step of $256$, which means $50$\% of overlap among consecutive windows. The proposed framework takes each data segment with shape ($512$,$8$,$16$) as input and outputs the predicted label among $66$ gestures.

\vspace{.05in}
\noindent
\textbf{3.2. Micro Path: Extraction of Microscopic Neural Drive Info.}

\noindent
Here, first, we present the fundamentals of the developed BSS algorithm for extracting MUSTs from HD-sEMG signals. Then we elaborate on how the $\MM$ framework is employed for classification of MUAPs as a by-product of the BSS method. Multi-channel sEMG signals are generated as a convolutive mixture of a set of impulse trains representing the discharge timings of multiple MUs, i.e.,
\begin{eqnarray}
\x_{i}(\t) =  \Sigma_{\l=0}^{\L-1} \Sigma_{j=1}^{\N} \h_{ij} (\l) \s_{j} (\t-\l) + \bm{\nu}_{i}(\t),\label{eq:gen}
\end{eqnarray}
where $\x_{i}(\t)$ is the $\i^{th}$ channel's EMG data (from the entire $\M$ channels); $\h_{ij} (\l)$ is the action potential of the $\j^{th}$ MU (from the entire $\N$ extracted MUs) measured at the $\i^{th}$ channel; $\s_{j} (\t)$ is the MUST at the $\j^{th}$ MU, and; $\bm{\nu}_{i}$ is the additive white noise at channel $\i$. Additionally, $\t$ is the time index; $\D$ is the duration of sEMG recordings; and $\L$ is the duration of MUAPs. Eq.~\eqref{eq:gen} is represented in matrix~form as 
\begin{eqnarray}
\X(\t) =  \Sigma_{\l=0}^{\L-1}  \H(\l) \S(\t-\l) + \underline{\bm{\nu}}(\t) \label{matrix},
\end{eqnarray}
where $\X(\t) \!=\! [{\x}_{1}(\t), {\x}_{2}(\t), \dots, {\x}_{\M}(\t)] ^ {\T}$ and
$\S(\t) \!=\! [{\s}_{1}(\t), {\s}_{2}(\t),$ $\dots, {\s}_{\N}(\t)] ^ {\T}$ are the recordings of all the $\M$ electrode channels and the MUSTs of all the $\N$ extracted sources at time $\t$, respectively. Term $\H(\l)$ is the ($\M\times\N$) matrix of action potentials, which is considered to be constant in duration $\D$. The convolutive equality of Eq.~\eqref{matrix} is the basic BSS assumption.

The objective is to find the maximum number of independent matrices $\S(\t)$ from Eq.~\eqref{matrix} if $\X(\t)$ is the only known parameter. Eq.~\eqref{matrix} can be written in an instantaneous form, where the source vectors are extended with their $\L-1$ delayed contributions. Additionally, to adapt the model for BSS conversions, the observation vectors are extended with their $\T$ delayed versions, resulting in the following  final convolutive model
\vspace{-.075in}
\begin{eqnarray}
\widetilde{\X}(\t) =  \widetilde{\H} \hspace{0.05cm} \widetilde{\S}(\t) + \widetilde{\underline{\bm{\nu}}}(\t) \label{extend},
\end{eqnarray}
where each of the $\widetilde{\X}(\t)$, $\widetilde{\H}$, $\widetilde{\S}(\t)$, and $\widetilde{\underline{\bm{\nu}}}(\t)$ are the extended versions of the observation, MUAPS, sources, and noise matrices, respectively. Among the existing BSS approaches~\cite{bss} suggested for HD-sEMG decomposition, gradient Convolution Kernel Compensation (gCKC)~\cite{holobara,holobarb} and fast Independent Component Analysis (fastICA)~\cite{fastica} are of great prominence and frequently used in the literature. To achieve better accuracy, the utilized BSS algorithm~\cite{bss} is a combination of gCKC~\cite{holobara,holobarb} and fastICA~\cite{fastica} algorithms. In the gCKC method, the MUSTs are estimated  using a linear Minimum Mean Square Error (MMSE) estimator as follows
\vspace{-.075in}
\begin{eqnarray}
\hat{\s}_{j} (\t) = \hat{\c}_{\s_{j}x} ^ {\T} \C_{xx} ^{-1} \x(\t), \label{lmmse}
\end{eqnarray}
in which $\hat{\s}_{j} (\t)$ is the estimate of the $\j^{th}$ MUST at time $\t$, $\hat{\c}_{\s_{j}x} \approx \E\{\x (t) {\s_{j}} ^ {\T} (\t)\}$ is approximation of the unknown cross-correlation vector between the MUSTs and the observations, and $\C_{xx} = \E\{\x (t) \x ^ {\T} (\t)\}$ is the correlation matrix of observations. Term $\E\{\cdot\}$ indicates the mathematical expectation. According to Eq.~\eqref{lmmse}, as $\hat{\c}_{\s_{j}x}$ is unknown, a blind estimation of MUSTs is iteratively found with gradient descent~\cite{holobara}. On the other hand, in the fastICA, the goal is to estimate separation vectors $\w$ such that
\vspace{-.075in}
\begin{eqnarray}
\hat{\s}_{\j}(\t) = \w_{\j} ^ {\T} (\k) \Z(\t), \label{eq:fICA}
\end{eqnarray}
where $\hat{\s}_{\j}$ is the $\j^{th}$ MUST; $\w_{\j}$ is the $\j^{th}$ separation vector; and $\Z$ is the whitened matrix of observations. The separation vectors are identified through the fixed-point optimization algorithm~\cite{fastica,bss}. Note that term $\k$ in Eq.~\eqref{eq:fICA} denotes the separation vector identifying fixed-point iterations.
To provide analogous inputs for the Macro and Micro ViT-based models, we extracted MUSTs for windowed HD-sEMG signals of shape ($512$,$8$,$16$), separately. The number of iterations for estimating a new source is set to $7$, therefore, a maximum of $7$ sources that are either sparse or uncorrelated at time lag zero are found for each window. The length of MUAPs ($\L$) that are computed from extracted MUSTs in the next step are assumed to be $20$ samples. As stated in~\cite{bss}, extension factor $\T$ in Eq.~\eqref{extend} multiplied by the number of sEMG channels should be greater than the number of extracted sources multiplied by the length of MUAPs. Furthermore, it is empirically shown that extension factors greater than $16$ have almost the same impact on the number and quality of extracted MUSTs. Therefore, we set extension factor to $20$ to be greater than $\frac{\N\times\L}{\M}$. The silhouette threshold determining which sources are of acceptable quality in each iteration is set to $0.92$.

\begin{figure}[t!]
\centering
\includegraphics[scale=0.3]{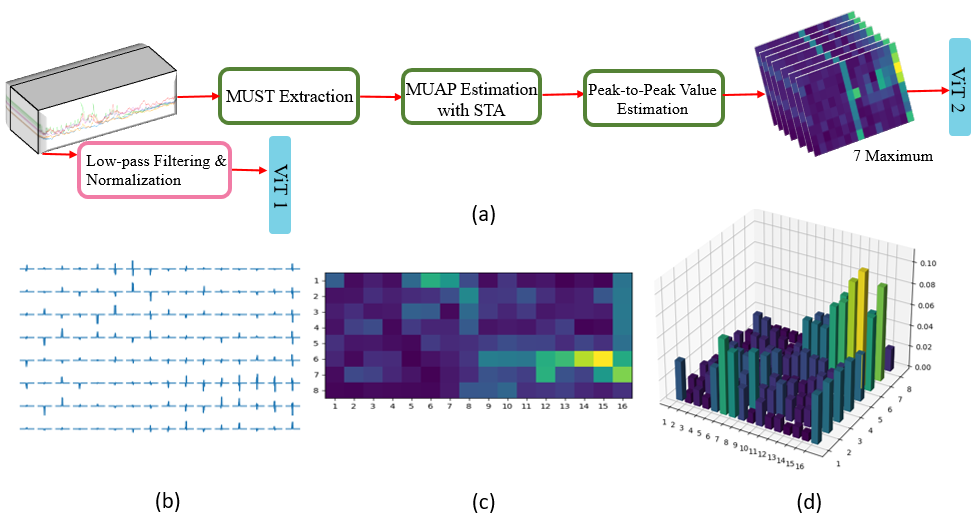}
\vspace{-0.45cm}
\caption{\footnotesize (a) Diagram of the adopted procedures for obtaining MUAP p-to-p images. (b) MUAPs for a single MU of the first windowed signal corresponding to the first repetition of gesture $1$ (bending the little finger). The MUAPs are estimated/shown for each channel separately. (c) p-to-p values of MUAPs represented as a 2D image. (d) 3D representation of MUAP p-to-p values. }
\label{sum-2d3d}
\end{figure}
\begin{table*}[t]
 \begin{center}
 \caption{\footnotesize Comparison of classification accuracy and STD for each fold and their average. The accuracy and STD for each fold is averaged over $19$ subjects.}
 \label{comp}
 {\begin{tabular}{c c c c c c c}
\hline
 \textbf{Model's Name}
& \textbf{Fold1(\%)}
& \textbf{Fold2}
& \textbf{Fold3}
& \textbf{Fold4}
& \textbf{Fold5}
& \textbf{Average}
\\
\hline

{Stand-alone Macro Model}
& 79.92 ($\pm$3.39)
& 91.43 ($\pm$2.48)
& 93.84 ($\pm$2.05)
& 92.57 ($\pm$2.28)
& 88.96 ($\pm$2.83)
& 89.34 ($\pm$2.61)
\\
{Stand-alone Micro Model}
& 81.53 ($\pm$3.45)
& 88.03 ($\pm$2.66)
& 89.63 ($\pm$2.39)
& 89.11 ($\pm$4.02)
& 84.92 ($\pm$2.97)
& 86.64 ($\pm$3.10)
\\
 The $\MM$
& \textbf{89.38 ($\pm$2.88)}
&\textbf{96.86 ($\pm$1.82)}
& \textbf{96.82 ($\pm$1.75)}
& \textbf{96.65 ($\pm$2.75)}
& \textbf{94.61 ($\pm$1.90)}
& \textbf{94.86 ($\pm$2.22)}
\\
\hline
\end{tabular}}
\end{center}
\vspace{-.4in}
\end{table*}

After extracting MUSTs for all the windows, the MUAP waveform for each channel and each extracted MU is calculated with the Spike-Triggered Averaging (STA) technique~\cite{dario_muap}. Following the work of~\cite{sta}, MUSTs are divided by a sliding window of $256$ samples with skip size of $256$, a MUAP of length $20$ is calculated for each window distinctly and the results are summed to give the final MUAP waveform. In such a way, $128$ ($8\times16$) MUAPs are acquired for each extracted source. In the final stage, the p-to-p values of each MUAP is calculated and fed to the Micro path's ViT-based architecture as a 2D ($8\times16$) image. This 2D image indicates the relative activation level of the neural drive under the area covered by the electrode grids, which is unique for each hand gesture. A summary of the adopted procedures from taking the raw HD-sEMG signals to calculating the MUAP p-to-p images is shown in Fig.~\ref{sum-2d3d}(a). Fig.~\ref{sum-2d3d}(b-d) illustrate the extracted MUAPs for a single MU of the first $512$-sample window of gesture $1$ (bending the little finger), 2D image of their p-to-p values, and a 3D representation of the p-to-p values, respectively. As can be seen, the muscles under the electrodes of the extensor grid were more active in the course of bending the little finger.

\vspace{.025in}
\noindent
\textbf{3.3. Combining the Two ViT-based Architectures}

\noindent
Before fusing the Macro and Micro neural information (extracted via the two ViT-based paths), each model is trained separately on its input dataset. Then, each model's weights are frozen and the FC layers at the end of each network that take the class token for classification are removed. Instead, the class tokens with shape of ($1\times\d$) and ($7\times\d$) for Macro network and Micro network, respectively, are concatenated and flattened. Here, $1$ represents the number of $512$-sample HD-sEMG signals, $7$ is the number of extracted MUs for each window, and $\d$ is the model's embedding dimension. The final feature vector has a dimension of $1,024$ ($8\times128$), which is fed to two consecutive FC trainable layers to form the classification results.

\vspace{-.15in}
\section{Experimental Results}  \label{sec:Sim}
\vspace{-.1in}
\begin{figure}[t!]
\centering
\vspace{-0.45cm}
\includegraphics[scale=0.33]{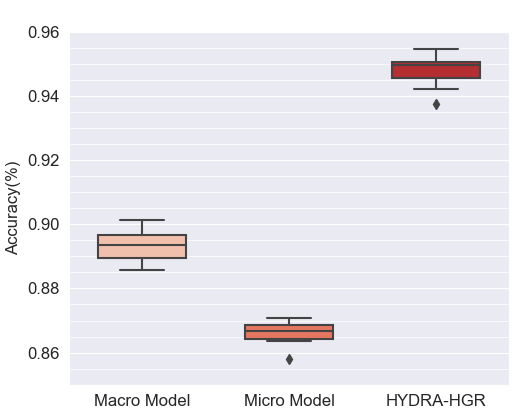}
\vspace{-0.45cm}
\caption{\footnotesize Boxplots and IQR of the $3$ models over all the $19$ subjects.}
\label{box}
\end{figure}
In this section, we evaluate classification accuracy and standard deviation (STD) of the Micro and Macro paths and the final fusion accuracy of the $\MM$. To facilitate a fair comparison between our study and future research works, we performed a $5$-fold cross validation, in which one repetition is opted out from training and used in testing each time. The mean accuracy among the participants is calculated for all the folds and their average is reported at the end. As one of the subjects in the dataset had incomplete information, all the studies are performed based on the recorded signals of $19$ subjects.
%
In the first experiment, windowed HD-sEMG signals are given to the Macro ViT-based architecture. In this architecture, the image size and the number of input channels are set to ($512\times16$) and $8$, respectively. The patch size is set to ($8\times16$) implying that there are $64$ temporal patches for each window. The model's embedding dimension ($\d$) is $128$ and the patches are divided into $8$ parallel heads in the transformer encoder. Adam optimizer is used with learning rate of $0.0001$ and weight decay of $0.001$. The network is trained with $20$ epochs with each batch containing $128$ data samples.
In the second experiment, 2D MUAP p-to-p images are fed to the Micro ViT-based architecture. In this architecture, the image size and the number of input channels are set to ($8\times16$), $1$ respectively. The patch size is set to ($8\times8$) implying that there are $2$ spatial patches for each p-to-p image, each corresponding to an ($8\times8$) electrode grid. The model's embedding dimension ($\d$) and number of heads is the same as its Macro counterpart. Similarly, Adam optimizer is used with learning rate of $0.0003$ and weight decay of $0.001$. The network is trained with $50$ epochs with each batch containing $64$ data samples. Each data sample, in this case, has a maximum of $7$ MUAP p-to-p images.
Finally, in the third experiment, both Macro and Micro ViT-based architectures are frozen, the feature vectors/matrices from each model are concatenated and fed to two trainable FC layers that map $1024$ features to $128$ and $128$ to $66$ which is the number of gestures in our study. This network is trained with $20$ epochs and a batch size of $128$ samples. Similarly, Adam optimizer is used with learning rate of $0.0005$ and weight decay of $0.0001$. A comparison of each model's accuracy and STD for each fold is represented in Table~\ref{comp}. The box plots showing accuracy and InterQuartile Range (IQR) measured for $19$ subjects is represented in Fig.~\ref{box} for each model. As can be seen in Table~\ref{comp}, the Macro Model's accuracy is, $\approx$ $3-4$\% better than that of the Micro Model for each fold. However, the $\MM$ improves the accuracy of each fold by at least $3$\% and the average accuracy of all the folds by $5.52$\%. Additionally, according to Fig.~\ref{box}, Micro Model has the least IQR and the $\MM$ stands significantly higher than the stand-alone models in terms of its accuracy among $19$ subjects.

\vspace{-.2in}
\section{Conclusion}\label{sec:Con}
\vspace{-.1in}
In this paper, a DNN-based model for classification of hand gestures from HD-sEMG signals is proposed in which two ViT-based architectures are fused to outperform the case of adopting each architecture singly. The training procedure in the proposed method comprises two stages; During the former, ViT-based architectures are trained separately on HD-sEMG signals and p-to-p values of the extracted MUAPs from raw HD-sEMG signals. In the latter, both models' weights are kept fixed and the feature sets obtained in their last layer are joined and given to two FC layers for final classification. We show that the classification accuracy of the fused framework is, on average, $5.52$\% and $8.22$\% greater than that of Macro and Micro Models, respectively.

\end{document}